\documentclass[a4paper, 11pt]{article}
\usepackage[hmargin=2.5cm,vmargin=3.5cm]{geometry}
\usepackage{geometry}
\usepackage[utf8]{inputenc}
\usepackage{amsmath}
\usepackage{mathrsfs}
\usepackage{graphicx}
\usepackage{subcaption}  
\usepackage{hyperref}
\usepackage{multicol}
\usepackage{mathtools}
\usepackage{eurosym}
\usepackage{pdfpages}
\usepackage[normalem]{ulem}
\usepackage{latexsym}
\usepackage{lmodern}
\usepackage{mathrsfs}
\usepackage{amssymb}
\usepackage[capposition=top]{floatrow}
\usepackage{lipsum}
\usepackage{setspace}
\usepackage{float}
\usepackage[english]{babel}
\usepackage{fancyhdr, afterpage}
\usepackage{bbm}
\usepackage{amsfonts}
\usepackage{threeparttable}
\usepackage{pdflscape}
\usepackage{tikz}
\usepackage{csquotes}
\usepackage[subnum]{cases}
\usepackage{booktabs}
\hypersetup{
colorlinks=true,
linkcolor=blue,
filecolor=black, 
urlcolor=blue,
citecolor=blue
}
\usepackage[authordate,bibencoding =auto,backend=biber,natbib, maxcitenames=1,uniquename=false,
  uniquelist=false, doi=false,isbn=false,url=false,eprint=false]{biblatex-chicago}
\renewcommand{\footnoterule}{%
  \kern -3pt
  \hrule width \textwidth height 1pt
  \kern 2pt
}
\title{Nowcasting distributions: a functional MIDAS model {\footnote{{Andrea Renzetti and Massimiliano Marcellino thank MUR-Prin 2022 - Prot. 20227YZ9JK, financed by the European Union - Next Generation EU, for partial financial support. We thank Vasco Botelho, Michael Pfarrhofer, Anna Simoni and Jonas Striakus  for helpful comments.}}} }
\author{Massimiliano Marcellino \thanks{Bocconi University and BAFFI-CAREFIN Center; email:  \href{mailto:massimiliano.marcellino@unibocconi.it}{massimiliano.marcellino@unibocconi.it}} \and Andrea Renzetti \thanks{Bocconi University and BAFFI-CAREFIN Center; email: \href{mailto:andrea.renzetti2@unibo.it}{andrea.renzetti2@unibo.it}.} \and Tommaso Tornese \thanks{Universitá Cattolica del Sacro Cuore, Milano; email: \href{mailto:tommaso.tornese@unicatt.it}{tommaso.tornese@unicatt.it}}}
\date{ This Draft: \today \\
}

\begin{document}
\maketitle
\begin{abstract}
We propose a functional MIDAS model to leverage high-frequency information for forecasting and nowcasting distributions observed at a lower frequency. We approximate the low-frequency distribution using Functional Principal Component Analysis and consider a group lasso spike-and-slab prior to identify the relevant predictors in the finite-dimensional SUR-MIDAS approximation of the functional MIDAS model. In our application, we use the model to nowcast the U.S. households' income distribution. Our findings indicate that the model enhances forecast accuracy for the entire target distribution and for key features of the distribution that signal changes in inequality.\bigskip \bigskip

\emph{J.E.L Classification Code: C32; E32 }

\bigskip

\bigskip
\end{abstract}
\clearpage

\section{Introduction }
Economists, policymakers, and social scientists have long recognized the profound implications of income distribution for social stability, economic growth, and overall well-being. The ongoing shifts in income distribution are central to understanding the evolution of economic inequality. Additionally, the income distribution influences aggregate demand through consumption and investments, as well as the demand for financial assets \citep{aiyagari1994uninsured,kaplan2018monetary,bayer2019precautionary,bilbiie2023inequality}. Therefore, closely monitoring the evolution of household income distribution is crucial, not only for tracking household income inequality per-se, but also for an accurate assessment of the current state of the economy.  Real-time monitoring of income distribution provides critical insights that can guide effective policy interventions and promote equality, economic development and stability. 

Most of the times, however, for what concerns the distribution of households' income we only have imperfect knowledge of the present state and even of the recent past.  Comprehensive surveys on households' income are often released with considerable lag, since gathering data from a large and diverse sample of households is time-consuming. Surveys often require in-person interviews, phone calls, or mailed questionnaires, which can take several months to complete.  This is especially true for large surveys that involve a representative sample of households and provide information on the annual stream of income, deriving not only from wages and salaries, but also from self employed income, social security benefits, interest and dividends, retirement income, unemployment compensations etc. 



The importance of monitoring real-time changes in income distribution necessitates seeking timely signals from various indicators available before the release of official data. In this work we introduce a functional MIDAS  (MIxed DAta Sampling) model for nowcasting the low frequency households' income distribution leveraging high frequency macroeconomic and financial indicators.\footnote{For an extensive treatment of MIDAS models see \citet{ghysels2025econometric}.} To exploit high-frequency macroeconomic indicators for nowcasting the households' income distribution, we face three main challenges. First, we need a finite dimensional approximation of the continuous distribution of household income. Second, we have to match the high frequency macroeconomic indicators with the low-frequency approximation of the distribution of income. Third, we might want to identify the relevant predictors from a potentially large number of macroeconomic and financial time series. These facts can easily cause a proliferation of parameters, potentially leading to overfitting with subsequent negative effects on forecast accuracy, particularly due to the small sample size resulting from annual or biannual observations in these types of household surveys. We address the first challenge using functional principal component analysis \citep{ramsay_silverman_2005}, which allows us to approximate the distribution of income with a relatively small number of basis functions. Approximating the distribution of income using functional principal components leads to a seemingly unrelated MIDAS regression (SUR-MIDAS) approximation of the functional MIDAS model. Still, when considering a possibly large number of high frequency indicators and a meaningful lag dynamics for these indicators, the SUR-MIDAS model is highly-parameterized. For this purpose, we extend the MIDAS group lasso spike and slab adaptive prior in \citet{MOGLIANI2021833} to a SUR framework. Since this prior is constructed to set exactly to zero the coefficients of the irrelevant predictors, this approach allows us to explore the usefulness of a potentially large number of predictors while performing variable selection. For the specification of the prior, we consider a re-parameterization of the original SUR model which enables an efficient equation by equation estimation of the parameters of the model.  Thus, the paper also contributes to the recent MIDAS literature by extending the sparse-group LASSO regression-based approach \citep{MLG2021,MOGLIANI2021833,babii2024nowcasting} to a SUR framework. 

Our model is a direct forecasting tool that matches high-frequency information from macroeconomic and financial indicators to predict the entire household income distribution before it is officially released.
In general, although it does not enable predictions for the micro-level distribution at a high frequency, the MIDAS framework allows leveraging high-frequency information for nowcasting the low-frequency micro-level income distribution. This applies both when the micro variable reports income for the fourth quarter of the current year and when they refer to the entire year, as in the application we consider. Indeed, since the MIDAS approach does not require filtering out missing observations of the micro-variable's low-frequency distribution, it avoids the issue of aggregating latent high-frequency distributions, as it would be required in the mixed frequency framework of \citet{schorfheide2015real}. This circumvents the need to impose additional assumptions when approximating the cross-sectional distributions. 

In the application, we use the model for nowcasting the distribution of households' income in the United States  from the Annual Social and Economic Supplement (ASEC) of the Current Population Survey (CPS). This annual survey on yearly household income is released in March of the following year. This data provides a comprehensive measure of the financial resources accrued by a household in a given year, and it is usually used for analyzing economic well-being and poverty status among different demographic groups in the United States. To investigate the  usefulness of high frequency macroeconomic indicators for nowcasting the distribution of household income, we set up a pseudo-real-time forecasting exercise. The set of predictors comprehends quarterly indicators concerning the evolution of aggregate income,  employment, financial accounts, government expenditure, consumption, interest rates and stock prices within the year. In the application, we find that leveraging quarterly macroeconomic and financial indicators with our model enhances forecast accuracy about the yearly cross-sectional distribution of household income  for the US. In particular, we find that exploiting high frequency information allows to enhance forecast accuracy of key features of the distribution that signal changes in inequality.

The rest of the article is organized as follows. In Section \ref{sec_funmidas} we introduce the functional MIDAS model and the group lasso-type prior for the parameters of its finite dimensional SUR-MIDAS approximation. To assess the small-sample performance of our modelling strategy and the prior specification, in section \ref{monte_carlo} we set-up a Monte Carlo simulation, which highlights the merits of our approach in realistic settings. Section \ref{application}
presents the application of the SUR-MIDAS to the problem of nowcasting the cross-sectional distribution of household income in the US released in the March ASEC of the CPS. Section \ref{conclusions} summarizes the main findings and concludes. The appendix presents additional theoretical and empirical results.

\section{A Functional-MIDAS model for nowcasting distributions}\label{sec_funmidas}

We analyze the dynamics of the cross-sectional distributions of income by modelling the associated Log-Quantile Density (LQD) function, i.e the first derivative of the quantile function corresponding to the distribution of interest \citep{parzen_1979,jones_1992}. That is:
\begin{equation}\label{transf}
    q_t(\tau) = log \left( \frac{\partial Q_t(\tau)}{\partial \tau} \right)
\end{equation}
where $Q_t(\tau)$ is the quantile function  mapping $\tau \in [0,1] \rightarrow [\underline{x}, \bar{x}]$ at each time $t$, for $t=1,\ldots, T$. As discussed by \cite{petersen} and \cite{Huber2024}, the use of the LQD function is convenient since $q_t(\tau)$  does not need to integrate to one or to obey non-negativity constraints, as it would be the case for the probability density function, or satisfy monotonicity constraints, which would be the case for the cumulated density function or the quantile function. As enforcing such constraints in a dynamic linear model is extemely challenging, modelling the LQD simplifies considerably the econometric analysis of the time variation in the distribution of interest. The explicit mapping from the probability density function \( f(x) \) into the LQD function, is defined by \( \psi_Q(f)(\tau) = -\log(f(Q(\tau))) \), where \( Q(\tau) \) is the quantile function. To model the dynamic interaction between the low frequency cross-sectional distribution of income and the high frequency macroeconomic indicators, we specify the following functional MIDAS regression model: 
\begin{equation}  \label{fun_midas}
        q_{t}(\tau) = c_q(\tau) + \sum_{l=0}^{p_x-1}\boldsymbol{B}_{qx,l}(\tau)L(l/m)\boldsymbol{x}_t^{(m)}  + \sum_{l=1}^{p_q}\int_0^1 B_{qq,l}(\tau,\tau')L(l)q_{t}(\tau')d\tau' + u_{q,t}(\tau)
\end{equation}
where $\boldsymbol{x}_{t}^{(m)}$ is the $n_x \times 1$ vector of high-frequency macroeconomic indicators, $L(.)$ is the lag operator, such that  $L(1/m)\boldsymbol{x}_t = \boldsymbol{x}_{t-1/m}^{(m)}$ and $L(1)q_t(\tau) = q_{t-1}(\tau)$. The function $q(\tau)$ and the high-frequency macroeconomic indicators $\boldsymbol{x}_t^{(m)}$ are assumed to be sampled at different frequencies. For instance, when the cross-sectional distribution is observed yearly and the macroeconomic indicators are observed quarterly, then $m = 4$.\footnote{Note that we are considering the same number of lags for all the high frequency macroeconomic indicators $n_x$. This is for ease of exposition.} Hence, in the model, the LQD function is expressed as a combination of $p_x$ lags of the high frequency macroeconomic indicators and its own $p_q$ lags. We assume that the LQD function admits the Karhunen-Loéve expansion
\begin{equation}\label{kl_theorem}
    q_t(\tau) = \mu(\tau) + \sum_{k=1}^{\infty}\sqrt{\lambda_k}z_{t,k}h_k(\tau)
\end{equation}
and we approximate $q_t(\tau)$ by truncating the infinite sum in (\ref{kl_theorem}) at a level $K$, namely:

\begin{equation}\label{finiteapprox}
     q_t(\tau) = \mu(\tau) + \sum_{i=1}^K h_i(\tau)f_{t,i} =  \mu(\tau)  +  \boldsymbol{h}_K(\tau)'\boldsymbol{f}_{t;K}
\end{equation}
where $\boldsymbol{h_K(\tau)}$ is a $K \times 1$ vector of basis functions, while $\boldsymbol{f}_{t;K}$ is a $K \times 1$ vector of coefficients or scores associated to the basis functions. Once the LQD function is approximated using $K$ basis functions, the functional MIDAS model in (\ref{fun_midas}) can be rewritten as a SUR-MIDAS (Seemingly Unrelated Regression - MIDAS) model for $\boldsymbol{f}_{t;K}$, that is 

\begin{equation}\label{surmidas}
    \boldsymbol{f}_{t;K} = \boldsymbol{\phi}_{0} + \sum_{l=0}^{p_x-1} \boldsymbol{\Phi_{fx,l}}L(l/m)\boldsymbol{x}_t^{(m)}  + \sum_{l = 1}^{p_q}\boldsymbol{\Phi_{ff,l}}L(l)\boldsymbol{f}_{t;K} + \boldsymbol{u_{f,t}}
\end{equation}
with $E[\boldsymbol{u_{f,t}}] = \boldsymbol{0}$ and $E[\boldsymbol{u_{f,t}}\boldsymbol{u_{f,t}}'] = \boldsymbol{\Omega}$.\footnote{Appendix \ref{approx_surmidas} reports the steps involved to go from the Functional MIDAS model to its SUR-MIDAS model representation.} It is easy to see that  when the number of basis functions in the approximation of the log-quantile-density function $K$ is big, the SUR-MIDAS model becomes highly-parameterized.  To approximate the LQD functions $q_t({\tau})$ using a relatively small number of basis functions, we resort to Functional Principal Component Analysis (FPCA, \cite{ramsay_silverman_2005}). While this approach allows to obtain a SUR-MIDAS model for a vector $\boldsymbol{f}_{t;K}$ of moderate dimensions, in our application we still face two challenges. First, we might consider incorporating a wide array of macroeconomic indicators and then select which of them are most relevant. Second, we might need a potentially large number of lags of the high frequency macroeconomic indicators to capture a meaningful lag dynamics. For example, in a typical application in which the cross-sectional distribution is observed at the yearly frequency and the macroeconomic indicators are observed at the quarterly frequency, including an extra macroeconomic indicator with the corresponding observations for just one year leads to $4K$ additional coefficients in the model. In general, since the model features $Kp_xn_x$
coefficients on the high frequency macroeconomic indicators and $K^2p_{q}$ coefficients on the lags of $\boldsymbol{f_{t;K}}$, the number of parameters increases as the number of high frequency indicators $n_x$ or the number of their lags $p_x$ increases, but also when the number of lags of the log-quantile-density function $p_q$ increases.  This proliferation of parameters can easily lead to overfitting since working with yearly cross-sectional income distributions typically implies small sample sizes in most applications. We address this issue by considering a group lasso spike and slab type prior for the coefficients in the MIDAS-SUR model \citep{xu_ghosh_2015,MOGLIANI2021833}. Section \ref{fpca} below, presents the details concerning the approximation of the LQD function with FPCA, while section \ref{lassoprior} presents the group lasso-type prior for the SUR-MIDAS model; section \ref{forecasting} discusses forecasting and nowcasting, and section \ref{estimation} introduces the estimation algorithm.

\subsection{Approximation of the LQD function by FPCA}\label{fpca}

Our estimation strategy follows a two-step approach. First, we approximate the low-frequency distributions using functional principal component analysis. Next, we estimate the SUR MIDAS model to generate nowcasts of the factors, and consequently, of the corresponding distributions. \footnote{Conditioning on the estimates of the eigenfunctions from FPCA in the first step, the factors can also be treated as random and estimated jointly with the parameters of the SUR MIDAS model in the Gibss Sampler by the Kalman filter \citep{carter1994gibbs} or by the precision sampler \citep{chan2009efficient}.} In this section, we detail the approximation of the targeted micro-variable distribution through FPCA. The basis function approximation in equation (\ref{finiteapprox}) allows to approximate non-parametrically potentially very flexible distributions. FPCA can be used to determine the ``optimal basis functions" for approximating functional data. These basis functions are derived to capture the maximum variance in the data with a given number of components, ensuring that the representation is both efficient and comprehensive. In the context of approximating distribution functions, \citet{petersen} has pointed out the advantages of exploiting the LQD transformation. The LQD transformation ensures that the original density functions are mapped into a linear Hilbert space, which is essential for the proper application of FPCA. This approach retains the interpretability of the principal components while overcoming the challenges posed by the nonlinear constraints intrinsic in the space of  density functions. Additionally, the LQD transformation leads to more efficient and meaningful representations of the variability in a sample of densities, as it captures both vertical and horizontal variations more effectively than direct FPCA on densities. Observations from the LQD functions $q_t(\tau)$ are obtained by evaluating 
\begin{equation}\label{eq:LQD}
    q(\tau) = -log(\hat{f}(Q(\tau)))
\end{equation}
on a set of grid points $\tau_{1}, \ldots, \tau_{N^{grid}}$. In Equation (\ref{eq:LQD}), $\hat{f}(.)$ is a kernel smoothed density estimate of the probability density function namely:
\begin{equation} 
\hat{f}(x) = \frac{1}{N^{grid} h} \sum_{i=1}^{n} K \left( \frac{x - x_i}{h} \right)
\end{equation}
We resort to FPCA in order to approximate $q_t(\tau)$
in equation (\ref{finiteapprox}). Specifically, the basis functions $\boldsymbol{h_K(\tau)}$ and the scores $\boldsymbol{f}_{t;K}$ are obtained by  static FPCA of the deviation of $q_t(\tau)$, observed over $t=1,\ldots, T$,  from their mean.  More precisely, to represent the continuous functions $q_t(\tau)$ in a finite dimensional data matrix $\boldsymbol{Q} \in \mathbb{R}^{N^{grid} \times T}$,  we center the data matrix by subtracting from each LQD function $q_t(\tau)$ evaluated on a grid $\tau=\tau_{1},...,\tau_{N_{grid}}$ a simple estimate of their mean obtained as $\hat{\mu}(\tau) = \frac{1}{T} \sum_{t=1}^{T}q_t(\tau)$, to get $\boldsymbol{Q}$. Then we apply Singular Value Decomposition (SVD) to the centered matrix, namely
\begin{equation}
    \boldsymbol{Q} = \boldsymbol{U S V^T}
\end{equation}
where $\boldsymbol{U}$ and $\boldsymbol{V}$ are orthogonal matrices, and $\boldsymbol{S}$ is a diagonal matrix containing the singular values (we omit the dependence on $\tau$ just for notation convenience). To obtain the basis functions \( \boldsymbol{h(\tau)} \), we select \( K \) columns from \( \boldsymbol{V} \), which represent the eigenbasis associated with the \( K \) largest eigenvalues. The $\boldsymbol{f}_{t;K}$ in (\ref{finiteapprox}) are therefore given by the principal component scores corresponding to the selected $K$ eigenbasis. \citet{petersen} demonstrates the consistency of these estimates and we refer to this paper for rates of convergence.

\subsection{A Group Lasso Shrinkage prior for the SUR-MIDAS model}\label{lassoprior}
Once the LQD functions are approximated by FPCA, the functional MIDAS model in (\ref{fun_midas}) can be rewritten as the SUR-MIDAS model for the principal component factors $\boldsymbol{f}_{t;K}$ in (\ref{surmidas}). As anticipated above, this model is highly parameterized. In this section we define a group lasso-type shrinkage prior for the coefficients in the SUR-MIDAS model.  
This prior is meant to  select the most relevant predictors and, at the same time, shrinking the coefficients within the selected groups to zero. More in detail, this prior extends the MIDAS adaptive group lasso prior of \citet{MOGLIANI2021833} to a SUR framework. We collect the lags of the high frequency macroeconomic indicators  and the lags of the low frequency factors in the vector $\boldsymbol{x}_t = [vec([ \boldsymbol{x}_{t-1/m}^{(m)} \ldots  \boldsymbol{x}_{t-p_x/m}^{(m)}]')',vec([\boldsymbol{f}_{t-1},\ldots,\boldsymbol{f}_{t-p_q}]')']'$. Then the we rewrite the model (\ref{surmidas}) as
\begin{equation}\label{compact_gen}
    \boldsymbol{f}_{t} = \boldsymbol{\Phi}'\boldsymbol{x}_t + \boldsymbol{u}_t  \hspace{1cm}\boldsymbol{u}_t  \sim\mathcal{N}(\boldsymbol{0}, \boldsymbol{\Omega})
\end{equation}
where we are setting $ h = 0$ (nowcasting) and consider $\boldsymbol{\phi}_0 = 0$ just for the sake of the exposition.\footnote{For ease of exposition we also suppress the $K$ subscript in $\boldsymbol{f}_{t;K}$ and just write it as $\boldsymbol{f}_{t}$ } In this notation, the matrix of coefficients $\boldsymbol{\Phi}$ of dimension $ (n_xp_x + Kp_q) \times K$ is storing the coefficients on both the lags of the high frequency indicators and the low frequency factors in all the $K$ equations of the SUR model. In equation (\ref{compact_gen}) we assume normality of the error term $\boldsymbol{u}_t$ and allow for possibly correlated errors across equations, meaning that $\mathbb{E}[u_{1t}^2] = \omega^2_{i}$ and $\mathbb{E}[u_{i}u_{k}]= \omega_{ik}$ for $k\neq i$. Rewriting the SUR model equation by equation we have
\begin{equation}
\begin{aligned}
    f_{1t} & = \boldsymbol{x_{t}}'\boldsymbol{\phi_1} + u_{1t}  \\
    f_{2t} & = \boldsymbol{x_{t}}'\boldsymbol{\phi_2} +  u_{2t} \\
    \vdots \\
    f_{Kt} & = \boldsymbol{x}_{t}'\boldsymbol{\phi_K}  + u_{Kt}   \\
\end{aligned}
\end{equation}
where the vector $\boldsymbol{\phi}_i$ for $i =1, \ldots, K$ is the $i^{th}$ column of $\boldsymbol{\Phi}$. We re-parameterize the SUR model following \citet{alsozellner2010} as follows

\begin{equation}\label{repa_model}
\begin{aligned}
    f_{1t} & = \boldsymbol{x_{t}}'\boldsymbol{\phi_1} + \varepsilon_{1t} \hspace{5cm}  \varepsilon_{1t}=u_{1t} \sim \mathcal{N}(0, \sigma^2_1) \\
    f_{2t} & = \boldsymbol{x_{t}}'\boldsymbol{\phi_2} + \alpha_{21} u_{1t} + \varepsilon_{2t}  \hspace{3.5cm}  \varepsilon_{2t} \sim \mathcal{N}(0, \sigma^2_2) \\
    \vdots \\
    f_{Kt} & = \boldsymbol{x}_{t}'\boldsymbol{\phi_K} + \alpha_{K,1} u_{1t} + \ldots + \alpha_{K,K-1} u_{K-1t} + \varepsilon_{Kt}  \hspace{2cm}  \varepsilon_{Kt} \sim \mathcal{N}(0, \sigma^2_K) \\
\end{aligned}
\end{equation}
This re-parameterization is convenient since it allows to rewrite the system of equations of the SUR model as a set of univariate regressions, with uncorrelated error terms $\varepsilon_{it}$. Clearly, the correlation across the equations remains evident, originating from the product of the $u_{i,t}$ terms and the coefficients $\alpha_{i,c}$ for $i = 2, \ldots, K$ and $c=1, \ldots, i-1$ . We specify a prior for the elements of  $\boldsymbol{\alpha_{i,.}} = [\alpha_{i,1}, \ldots, \alpha_{i i-1}]$ and for the variances $\sigma^2_1, \ldots, \sigma^2_K$ which implies a symmetric prior for the variance covariance matrix of the error term in the original model $\boldsymbol{\Omega}$.  More specifically, we assume
\begin{equation}
    (\alpha_{i,c}| \sigma_{i}^2) \sim \mathcal{N} \left(0, \frac{\sigma^2_{i}}{s_c^2} \right) \hspace{1.5cm} 1\leq c \leq i = 2, \ldots, K
\end{equation}
\begin{equation}
    \sigma_{i}^2 \sim \mathcal{IG}\left(\frac{v_0 + i - K}{2}, \frac{s_i^2}{2}\right) \hspace{1cm} i = 1, \ldots, K 
\end{equation}
It can be shown that this specification implies an \textit{Inverse-Wishart} prior for the variance covariance matrix of the error term in the original model (\ref{surmidas}), namely\footnote{The proof of this result can be found in \citet{JCC_QE}. }
\begin{equation}
\boldsymbol{\Omega} \equiv \mathbb{E}[\boldsymbol{u}_t \boldsymbol{u}_t'] = \boldsymbol{\tilde{A}}^{-1} \boldsymbol{\Sigma} \left(\boldsymbol{\tilde{A}}^{-1}\right)' \sim \mathcal{IW}(\boldsymbol{S}, v_0)
\end{equation}
where $\boldsymbol{\Sigma} = diag(\sigma_{1}^{2},\ldots,\sigma_{K}^{2})$, $\boldsymbol{\tilde{A}}$ is the lower triangular matrix with unit elements on the main diagonal and rows given by  $\boldsymbol{\tilde{\alpha}_{i,.}} = [-\alpha_{i,1}, \ldots, -\alpha_{i i-1}]$   and $\boldsymbol{S} = diag(s_1^2, \ldots, s_K^2)$. Hence, this specification keeps the symmetry of the prior distribution for the variance-covariance matrix of the error term in the original model. At the same time, the specification allows to directly specify a prior for the coefficients in $\boldsymbol{\Phi}$. In this respect, our SUR-MIDAS model features an unrestricted lag-dynamics for both the high frequency macroeconomic indicators and for the low frequency factors. To specify a group lasso-type prior for the elements in $\boldsymbol{\phi_i}$ i.e. the coefficients in the $i=1,\ldots,K$ equations of the SUR-MIDAS model, the natural approach would be to define $n_x$ groups, one for each high frequency macroeconomic indicator, and $K$ groups, one for each low frequency factor. This makes a total of $G = n_x + K$ groups with $n_x$ groups  of dimension $p_{x}$
(the number of lags for each high frequency regressor) and $K$ groups  of dimension $p_{q}$ (the number of lags for each low frequency factor). 
Alternatively, as in \citet{MOGLIANI2021833}, one can reduce the number of unknown parameters by restricting the lag-dynamics of the high frequency macroeconomic indicators $\sum_{l=0}^{p_x-1}L(l/m)\boldsymbol{x}_t^{(m)}$ and low-frequency factors $\sum_{l=1}^{p_q}L(l)\boldsymbol{f}_{t;K}$, exploiting the Almon lag polynomial. More specifically, the direct representation of the Almon lag polynomial allows to keep the model linear in the parameters, by just considering a transformation of the regressors $\boldsymbol{x}_{t}$. In particular, considering an Almon lag polynomial leads to the following transformation of the regressors
\begin{equation}
  \boldsymbol{z_{j,t}} = \boldsymbol{W}\boldsymbol{x}_{j,t}
\end{equation}
where $\boldsymbol{x_{j,t}} = [x_{jt-1/m}^{(m)}, \ldots, x_{j,t-p_x/m}^{(m)} ]'$ for the lags of the high frequency variables and $\boldsymbol{x_{j,t}} = [x_{jt-1}, \ldots, x_{j,t-p_q}]'$ for the lags of the low frequency factors and $\boldsymbol{W}$ is the polynomial weighting matrix of this transformation. This approach allows to further reduce the dimension of the groups made of the lags of the high frequency indicators and of the lags of the low frequency factors to $p_a + 1 - r_a$ where $p_a$ is the order of the Almon lag  polynomial and $r_a$ are the number of restrictions on the shape of the Almon lag polynomial.\footnote{The weighting matrix $\boldsymbol{Q}$ is of dimension $(p_a + 1 - r_a) \times p_x$ for the transformation of the high frequency variables and of dimension  $(p_a + 1 - r_a) \times p_q$ for the transformation of the low frequency factors. Note that just for ease of the exposition we are considering the same degree of the Almon lag polynomial for all the groups. Clearly, it is also possible to keep an unrestricted lag dynamics on the lags of the low-frequency factors  by just considering a linear transformation where $\boldsymbol{Q} = \boldsymbol{I}_{p_q}$ such that the dimension of the groups of each low frequency factor remains  $p_q$.} Exploiting the direct representation of the Almon lag polynomial, we rewrite the SUR-MIDAS model as:

\begin{equation}\label{repa_model2}
\begin{aligned}
    f_{1t} & = \boldsymbol{z}_t'\boldsymbol{\theta}_1 + \varepsilon_{1t} \hspace{5cm}  \varepsilon_{1t} \sim \mathcal{N}(0, \sigma^2_1) \\
    f_{2t} & = \boldsymbol{z}_t'\boldsymbol{\theta}_2 + \alpha_{21} u_{1t} + \varepsilon_{2t}  \hspace{3.5cm}  \varepsilon_{2t} \sim \mathcal{N}(0, \sigma^2_2) \\
    \vdots \\
    f_{Kt} & = \boldsymbol{z}_t'\boldsymbol{\theta}_K + \alpha_{K,1} u_{1t} + \ldots + \alpha_{K,K-1} u_{K-1t} + \varepsilon_{Kt}  \hspace{2cm}  \varepsilon_{Kt} \sim \mathcal{N}(0, \sigma^2_K) \\
\end{aligned}
\end{equation} 
where the vector $\boldsymbol{\theta}_i = [\boldsymbol{\theta_{i,1}}', \ldots, \boldsymbol{\theta_{i,G}}' ]'$ collects all the parameters of the Almon lag polynomial of each of the  $G = (n_x + K)$ groups in the $i^{th}$ equation of the SUR model. The vector $\boldsymbol{z}_t$ is instead storing the transformed regressors. Note that each vector $\boldsymbol{\theta}_{ij}$ is of dimension $(p_a + 1 - r_a) \times 1$ as, for ease of exposition, we are considering the same Almon lag polynomial order and the same number of restrictions for all the groups. Once we have obtained the groups and defined $\boldsymbol{\theta}_{ij}$ as the vector of coefficients in the $i^{th}$ equation
for the $j^{th}$ group  we specify the following prior \citep{xu_ghosh_2015,MOGLIANI2021833}
\begin{equation}
p(\boldsymbol{\theta_{ij}}|\sigma_i^2,\tau_{ij}^2,\pi_{i0}) = (1 - \pi_{i0}) \mathcal{N}(\boldsymbol{0}, \sigma^2\tau_{ij}^2 \boldsymbol{I_{g_{ij}}}) + \pi_{i0}\delta_{0}(\boldsymbol{\theta_{ij}})
\end{equation}
\begin{equation}\label{PRIOR_THETA}
p(\tau_{ij}^2) \sim Gamma \left( \frac{g_{ij}+1}{2}, \frac{\lambda_{ij}^2}{2} \right)
\end{equation}
\begin{equation}
 \pi_{i0}\sim Beta\left( c,d\right)   \end{equation}
for $i = 1, \ldots, K$ and $j=1,\ldots, G$. This prior introduces two types of shrinkage effects: a point mass at zero, resulting in exact zero coefficients (the spike component), and a Group Lasso prior applied to the slab component. Therefore this prior enhances variable selection at the group level while simultaneously shrinking the coefficients within the selected groups. 
\subsection{Forecasting } \label{forecasting}The MIDAS regression is a direct forecasting tool which results in different forecasting models for each forecast horizon \citep{ghysels_2004}. Collecting the parameters of the Almon lag polynomial for all the groups in the $i^{th}$ equation of the SUR-MIDAS model in $\boldsymbol{\theta}_i = [\boldsymbol{\theta_{i,1}}', \ldots, \boldsymbol{\theta_{i,G}}' ]'$ and defining $\boldsymbol{\Theta}$ as the matrix stacking the vectors $\boldsymbol{\theta}_i$ for $i = 1, \ldots, K$ on its columns, we can write the \textit{h-step} ahead direct forecast of the low frequency factors as
\begin{equation}
    \boldsymbol{f}_{t+h} = \boldsymbol{\Theta}'\boldsymbol{z}_t + \boldsymbol{u}_{t+ K} \hspace{2cm}  \boldsymbol{u}_{t+ K} \sim \mathcal{N}(\boldsymbol{0}, \boldsymbol{\Omega})
\end{equation}
The \textit{h-step} ahead direct forecast of the LQD function is therefore  
\begin{equation}
    q_{t+h}(\tau) = \mu(\tau) + \boldsymbol{h(\tau)}'\boldsymbol{f}_{t+h}.
\end{equation}
From the forecast of the LQD function we obtain the corresponding distribution function, as in \citet{kokoszka_2019}. More in detail, the backward mapping recovers the  \textit{h-step} ahead predicted density function from the log-quantile density function through \( f(x) = \exp\{-\psi_Q(f)(F(x))\} \). In this regard, our MIDAS approach allows to recover the distribution function regardless of whether the micro-variable is a flow or a stock. Unlike the standard mixed-frequency literature \citep{schorfheide2015real}, we do not treat the high-frequency distribution of the micro-level variable as latent. Therefore, we avoid the issue of aggregating missing high-frequency distributions, which can be non-trivial task, as noted by \citet{schorchang2022effects}. For instance, in the case of flow variables, our method can accommodate both low-frequency observations relative to the last high-frequency period and those spanning the entire low-frequency period. In other words, when predicting the household-level income distribution observed at a yearly frequency using quarterly predictors, our approach works regardless of whether the micro variable reports income for the fourth quarter of the current year or for the entire year. In the latter case, as with the income variable from the CPS ASEC in our application, specifying an aggregation rule for the missing quarterly distribution within the mixed-frequency framework of \cite{schorfheide2015real} would be complex and might require a more structured approach. This complexity arises because the observed yearly income distribution is a convolution of four unobserved quarterly income distributions and would require imposing additional structure by means of an appropriate aggregation rule.\footnote{\citet{DePolis2024} recently consider a mixed frequency functional VAR, treating the high frequency cross-sectional distributions as missing and imposing such intertemporal restrictions in the aggregation rule.} Hence, although it does not enable high-frequency predictions for the micro-level distribution, the MIDAS framework allows to impose less structure when approximating low frequency distributions and leverage high frequency information.  

\subsection{Bayesian Inference}\label{estimation}
As anticipated above, inference proceeds in two steps. First we approximate the LQD function using FPCA and obtain an estimate of the factors  $\boldsymbol{f}_{t}$. Then we estimate the SUR-MIDAS model 
using a Gibbs sampler that iteratively samples from the conditional posterior distributions of the parameters in model (\ref{repa_model2}), namely:

\begin{enumerate}
    \item Sample from $p(\boldsymbol{\theta}_{ij}|.)$ for $i = 1, \ldots, K$ and $j = 1, \ldots, G$.
    \item Sample from $p(\boldsymbol{\alpha_{i,.}}|.)$ for $i =2,\ldots, K$
    \item Sample from $p(\sigma_i^2|.)$ for $i =1,\ldots, K$
    \item Sample from $p(\tau_{ij}|.)$ for $i = 1, \ldots, K$ and $j = 1, \ldots, G$
    \item Sample $p(\pi_{i0}|.)$  for $i =1,\ldots, K$
\end{enumerate}
As anticipated above, the specification of the priors on the re-parameterized version of the model (\ref{repa_model})  enables the use of an efficient posterior sampling algorithm to produce inference about coefficients of the model equation by equation. In particular, to sample the coefficients in Step 1 we exploit the triangular algorithm proposed in \citet{carriero2019large} and corrected in  \citet{corrCARRIERO2022}. The formulas of the conditional posterior distributions are reported in the appendix \ref{condpost}. To set the hyper-parameter $\lambda^2_{ij}$, we follow the approach in \citet{MOGLIANI2021833} which is an Empirical Bayes approach that
relies on the so-called internal adaptive MCMC algorithms in \citet{atchade_2011}.\footnote{Alternatively we also report the formula of the conditional posterior of $\lambda^2_{ij}$ when treating it as random variable with Gamma prior distributions }


\section{Monte Carlo simulation}\label{monte_carlo}
We perform a Monte Carlo simulation to investigate the gains of exploiting high frequency variables for nowcasting a distribution observed at a lower frequency with our approach. In particular, we assume that the true LQD function of the low frequency distribution we aim to forecast has the following representation
\begin{equation}\label{finiteapprox_MC}
     q_t(\tau) = \mu(\tau) + \sum_{i=1}^3 h_i(\tau)f_{t,i} =  \mu(\tau) + \boldsymbol{h}(\tau)'\boldsymbol{f}_{t}
\end{equation}
We assume that the probability density function associated to this LQD function is defined on the support $(0,10)$. We consider the following SUR MIDAS dynamics for the elements of the vector of factors $\boldsymbol{f}_t$ 

\begin{equation}\label{dyn_LLF}
f_{it} = \alpha_i + \sum_{j=1}^{30} \beta_{ij} \sum_{c=0}^{24-1} \tilde{B}(c; \theta) L^{(m)}_c x_{j,t-h}^{(m)} + \sum_{l=1}^{2}\boldsymbol{\phi_{il}}'\boldsymbol{f}_{t-l} + u_{i,t}  
\end{equation}

\begin{equation}\label{dyn_HF}
x_{j,t}^{(m)} = \mu + \rho x_{j,t-1/m}^{(m)} + \epsilon_{j,t}    
\end{equation}
where $i=1, \ldots, 3$ is the number of factors and therefore of equations in the SUR-MIDAS representation of the functional MIDAS model. $j=1, \ldots, 30$ is the number of high frequency variables while $24$ is the lag length of the high frequency variables and $2$ is the number of lags for the low frequency factors. The high frequency variables are sampled at frequency $m = 4$, to mimic yearly low frequency observations and quarterly high frequency observations as in our application. The lag-polynomial $\tilde{B}(c; \theta)$ denotes the normalized weights (i.e., summing up to 1).  To set the weighting scheme we consider the DGP 1 in \citet{MOGLIANI2021833}. The lag dynamics of the low frequency factors is instead left unrestricted. The error terms $\boldsymbol{u}_t := (u_{1,t}, \ldots, u_{K,t})'$ and $\boldsymbol{\epsilon_t} := (\epsilon_{1,t}, \ldots, \epsilon_{n_x,t})'$ are i.i.d. with distribution
\begin{equation}
\begin{pmatrix}
\boldsymbol{u}_{t} \\
\boldsymbol{\epsilon}_t
\end{pmatrix} \sim \mathcal{N} \left( 
\begin{bmatrix}
\boldsymbol{0} \\
\boldsymbol{0}
\end{bmatrix}, 
\begin{bmatrix}
\boldsymbol{\Omega} & \boldsymbol{0} \\
\boldsymbol{0} & \boldsymbol{\Sigma_\epsilon} 
\end{bmatrix}
\right),
\end{equation}
Details on the other parameters of the DGP are in the Appendix 
\ref{sec:appendix_monte_carlo}. The basis functions in (\ref{finiteapprox_MC}) are selected as the eigenbasis corresponding to the three largest eigenvalues obtained through FPCA on the LQD function of a collection of \textit{Skew-t} distributions \citep{AC2003}. Each distribution is characterized by different location, shape, scale, tail thickness and it is truncated on the support (0,10).  

\begin{figure}[H]
    \centering
    \begin{subfigure}[b]{0.49\textwidth}
        \centering
    \includegraphics[width=\textwidth]{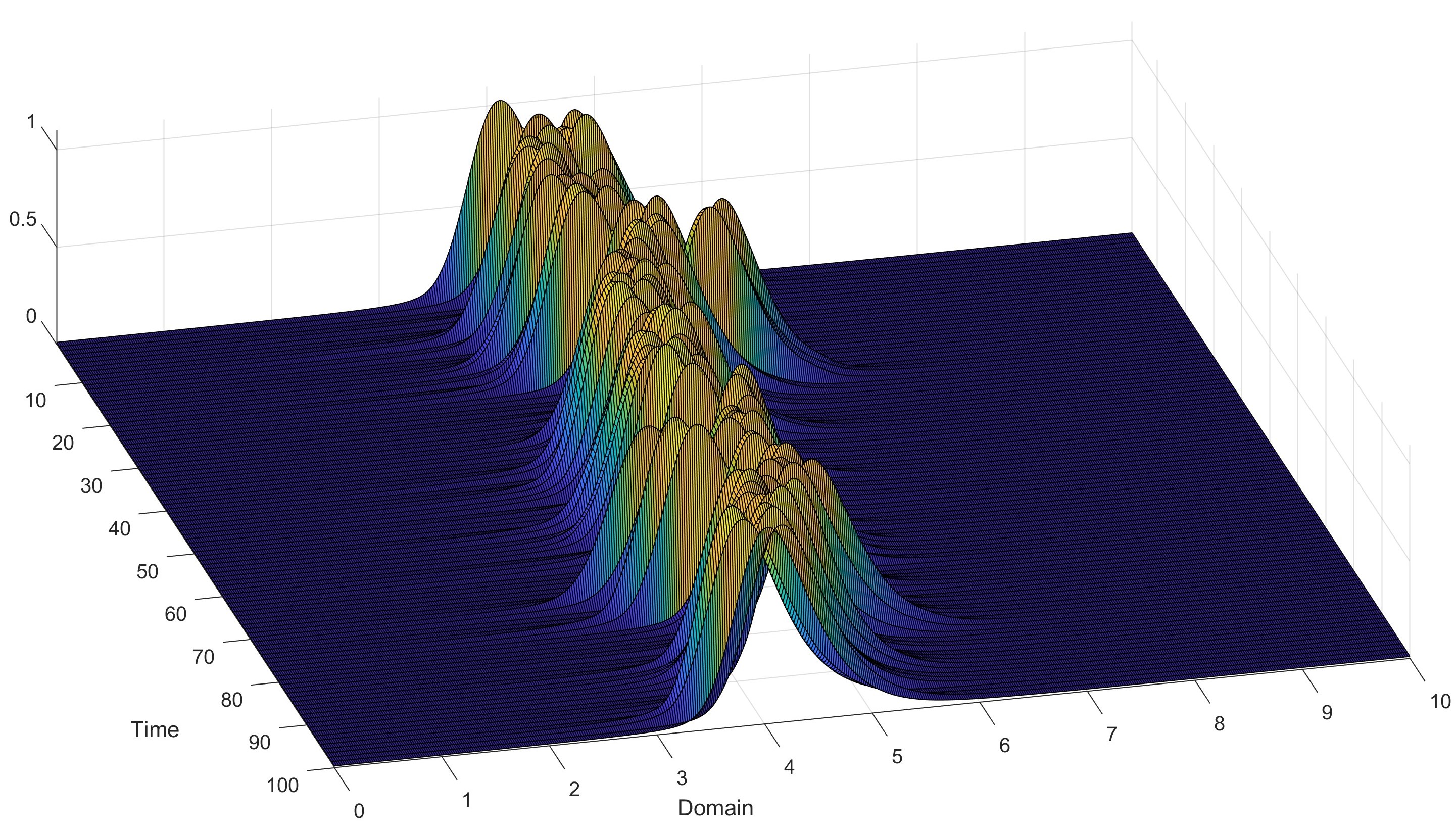}
        \caption{In-sample simulated distributions}
        \label{fig:left_image}
    \end{subfigure}
    \hfill
    \begin{subfigure}[b]{0.49\textwidth}
        \centering
    \includegraphics[width=\textwidth]{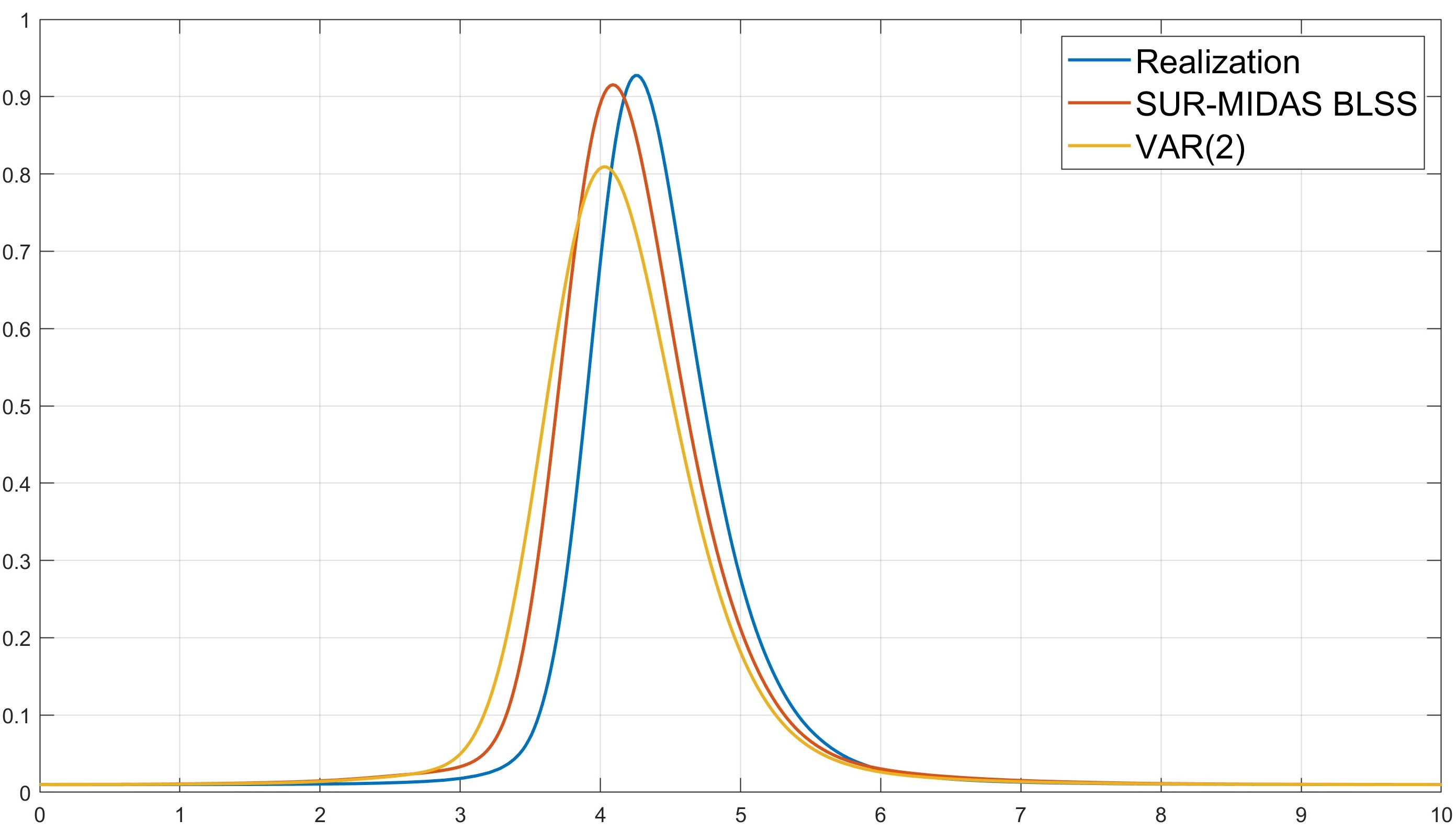}
        \caption{Out-of-sample true and predicted distribution}
        \label{fig:right_image}
    \end{subfigure}
    \caption{Output from one replication of the Monte Carlo }
    \label{fig:combined_images}
\end{figure}

As we want to assess whether our approach allows to correctly select the relevant high frequency predictors among the thirty high frequency indicators, we assume a sparse structure for the $\beta_{ij}$ in (\ref{dyn_LLF}). In other words, 
in each equation of the SUR-MIDAS model, only a subset of predictors is relevant for predicting the low-frequency factors and therefore changes in the distribution over time. We consider two sample sizes $T = \{ 60, 120 \}$ and the number of the Monte Carlo simulations is 500.  Figure \ref{fig:combined_images}, panel a)  shows the time series of the simulated distributions in one replication of the Monte Carlo. Figure \ref{fig:combined_images} panel b) shows the first out-of-sample realization of the true distribution, with the nowcast obtained from our functional VAR(2) model and a functional MIDAS model with LASSO type prior discussed in the text and a ridge type prior discussed in the appendix \ref{sec:ridge_type}. The point forecast of the distribution function is obtained by averaging across the posterior distribution draws. For ease of analysis we assume $h = 0$, i.e. a nowcasting model with high-frequency information fully matching the low frequency. While the SUR-MIDAS models leverage information from both the high frequency variables and the lags of the low frequency factors, the VAR(2) is just exploiting the information from the lags of the low frequency factors. 
To assess the accuracy of the nowcasts across the Monte Carlo replications, we compare the Kullback-Leibler (KL) distance and the Hellinger Distance (HD) to the true probability density function, the quantile scores (QS) and the root mean squared error (RMSE) of some selected moments of the distribution such as mean, variance, skewness and kurtosis. As a measure of dispersion we consider the interquartile range while as a measure of inequality we consider the Gini and the coefficient of variation (CV). 

Table \ref{tab:mc_results} reports the results from the Monte Carlo simulation. Even in small sample size $T= 60$, the spike and slab prior correctly selects the relevant predictors in the model, outperforming both the SUR-MIDAS with ridge type prior and the VAR(2). The Monte Carlo simulation shows that exploiting high frequency information allows to improve the forecasts from the entire distribution and of key features that signal changes in inequality such as the Gini Index and the coefficient of variation. 
\begin{table}[H]
    \centering
    \scalebox{0.75}{
    \begin{tabular}{lccccccc}
        \toprule
        \multicolumn{8}{c}{T=60} \\ 
        \bottomrule\toprule
        & Avg KL & Avg HD & Avg QS5 & Avg QS20 & Avg QS50 & Avg QS80 & Avg QS95 \\
        \midrule
        VAR(2) & 0.1651 & 0.1817 & 0.0760 & 0.0992 & 0.1195 & 0.1221 & 0.5795 \\
        RIDGE SUR-MIDAS  & 0.1204 & 0.1439 & 0.0713 & 0.0826 & 0.0934 & 0.1038 & 0.4422 \\
        BLASSO SUR-MIDAS  & 0.1029 & 0.1343 & 0.0645 & 0.0757 & 0.0861 & 0.0935 & 0.4188 \\
        \bottomrule
        \toprule
        & RMSE Mean & RMSE Variance & RMSE Skewness & RMSE Kurtosis & RMSE IR & RMSE GINI & RMSE CV \\
        \midrule
        VAR(2) & 0.1794 & 2.1970e-1 & 0.4830 & 2.4652 & 0.1129 & 0.0623 & 2.1684 \\
        RIDGE SUR-MIDAS & 0.1648 & 1.6270e-1 & 0.3840 & 1.8103 & 0.0438 & 0.0437 & 1.7081 \\
        BLASSO SUR-MIDAS & 0.1570 & 1.5640e-1 & 0.3908 & 1.8028 & 0.0492 & 0.0436 & 1.5925 \\
        \bottomrule
        \toprule
        \multicolumn{8}{c}{T=120} \\ 
        \bottomrule\toprule
        & Avg KL & Avg HD & Avg QS5 & Avg QS20 & Avg QS50 & Avg QS80 & Avg QS95 \\
        \midrule
        VAR(2) & 0.1685 & 0.1794 & 0.0781 & 0.0995 & 0.1193 & 0.1247 & 0.5755 \\
        RIDGE SUR-MIDAS & 0.1137 & 0.1393 & 0.0792 & 0.0846 & 0.0904 & 0.0975 & 0.4035 \\
        BLASSO SUR-MIDAS & 0.0886 & 0.1208 & 0.0725 & 0.0740 & 0.0777 & 0.0838 & 0.3476 \\
        \bottomrule
        \toprule
        & RMSE Mean & RMSE Variance & RMSE Skewness & RMSE Kurtosis & RMSE IR & RMSE GINI & RMSE CV \\
        \midrule
        VAR(2) & 0.1795 & 2.1950e-1 & 0.4864 & 2.4725 & 0.0999 & 0.0631 & 2.1946 \\
        RIDGE SUR-MIDAS & 0.1633 & 1.5130e-1 & 0.3969 & 1.7699 & 0.0414 & 0.0432 & 1.6184 \\
        BLASSO SUR-MIDAS & 0.1383 & 1.3680e-1 & 0.3453 & 1.6040 & 0.0319 & 0.0384 & 1.4008 \\
        \bottomrule
    \end{tabular}
 }
    \caption{Forecast accuracy in the Monte Carlo Simulation}
    \label{tab:mc_results}
    \vspace{0.25cm}\hspace{0cm}\parbox{1.2\textwidth}{\scriptsize Notes: The table reports the results from Monte Carlo simulations with $T=60$ in the first two sections and $T=120$ in the \\ last two sections. All values are rounded to four decimal places. Very small and very large numbers are written \\using scientific notation for readability.}
\end{table}

\section{Nowcasting household income distribution in the US}\label{application}

\subsection{Data and design of the nowcasting exercise}
We now use our functional MIDAS model
to nowcast the yearly cross-sectional distribution of households' income in the United States. We focus on the variable \texttt{hhincome} from the Annual Social and Economic Supplement (ASEC) of the Current Population Survey (CPS). This variable aggregates the income received by all members of a household from various sources over the past year, including wages and salaries, self-employment earnings, interest, dividends, rents, and other forms of income like social security benefits, retirement pensions, and public assistance. In general, this variable provides a comprehensive measure of the financial resources available to a household, and it is typically used for analyzing economic well-being and poverty status among different demographic groups in the US. This annual survey on yearly household income is released in March of the following year. Throughout the year, up until the release date, information about the latest developments in both the real and the financial markets gradually becomes available. This data includes macroeconomic and financial indicators, which might be especially valuable for producing a more precise forecast of the yearly income distribution before its official announcement in March. To leverage this information effectively, we include these high frequency indicators in a functional MIDAS model for the distribution of income. The set of quarterly macroeconomic indicators and their transformation can be found in table \ref{tab:variables}. The household-level nominal income given by the variable \texttt{hhincome} is normalized by dividing by 2/3 of nominal GDP per-capita and then transformed by computing the inverse hyperbolic sine function as in \citep{schorchang2021heterogeneity}.\footnote{The inverse hyperbolic sine transformation is given by \[
x = g(z) = \ln\left( z + \sqrt{ z^2 + 1}\right) = \sinh^{-1}(z)
\]} 
\begin{table}[htp]
\caption{\label{tab:variables} List of quarterly indicators}
\scalebox{0.7}{
\begin{tabular}{llll}
\hline
\textbf{Variable} & \textbf{Mnemonic} & \textbf{Transformation} \\
\hline\hline
Real Gross Domestic Product & GDPC1 & $\% \Delta$ \\
Real Government Receipts & FGRECPTx & $\% \Delta$ \\ 
Federal Government: Current Expenditures & FGEXPND & $\% \Delta$ \\
Real Personal Consumption Expenditures & PCECC96 & $\% \Delta$ \\
Real Disposable Personal Income & DPIC96 & $\% \Delta$ \\
Real Exports of Goods & EXPGSC1 & $\% \Delta$ \\
Real Imports of Goods \& Services & IMPGSC1 & $\% \Delta$ \\
All Employees: Total nonfarm & PAYEMS & $\% \Delta$ \\
Civilian Labor Force Participation Rate & CIVPART & $\% \Delta$ \\
Civilian Unemployment Rate & UNRATE & $\% \Delta$ \\
Number of Civilians Unemployed for 27 Weeks and Over & UEMP27OV & $\% \Delta$ \\
Average Weekly Hours Of Production And Nonsupervisory Employees & AWHNONAG & $\% \Delta$ \\
Average (Mean) Duration of Unemployment & UEMPMEAN & $\Delta$ \\
Housing Starts & HOUST & $\% \Delta$ \\
Real Average Hourly Earnings of Production and Nonsupervisory Employees: Total & AHETPIx & $\% \Delta$ \\
Real Average Hourly Earnings of Production and Nonsupervisory Employees Construction & CES2000000008x & $\% \Delta$ \\
Real Average Hourly Earnings of Production and Nonsupervisory Employees Manufacturing & CES3000000008x & $\% \Delta$ \\
Consumer Price Index all items & CPIAUCSL  & $\% \Delta$ \\
S\&P 500  & SP\_500 & $\% \Delta$ \\
Effective Federal Funds Rate & FEDFUNDS & $\Delta$ \\
10-Year Treasury Constant Maturity Rate (Percent) & GS10 & $\Delta$ \\
Cons. Expectations & UMCSENTx & level \\
\hline
\end{tabular}}

    \vspace{0.25cm}\hspace{0cm}\parbox{1.2\textwidth}{\scriptsize Notes: The symbols refer to the following transformations $\Delta$: first difference, $\% \Delta$: growth rate.   }
\end{table}

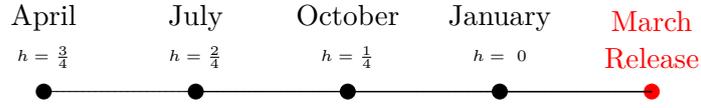
\begin{figure}[H]
\caption{Updates in the nowcasting exercise}\label{fig:updates}
    \begin{tikzpicture}
    \coordinate (A) at (0,0);
    \coordinate (B) at (2,0);
    \coordinate (C) at (4,0);
    \coordinate (D) at (6,0);
    \coordinate (E) at (8,0);
    
    \draw[dotted, gray] (A) -- (E);

    \fill[black] (A) circle (3pt);
    \fill[black] (B) circle (3pt);
    \fill[black] (C) circle (3pt);
    \fill[black] (D) circle (3pt);
    \fill[red] (E) circle (3pt);
    \draw[red, thick] (E) node[above=5pt, align=center, color=red] {March \\ Release}  ;

    \node[above=5pt, align=center] at (A) {April \\ \tiny{$h =\frac{3}{4}$}};
    \node[above=5pt, align=center] at (B) {July \\ \tiny{$h =\frac{2}{4}$}};
    \node[above=5pt, align=center] at (C) {October \\ \tiny{$h =\frac{1}{4}$}};
    \node[above=5pt, align=center] at (D) {January \\ \tiny{$h = \frac{}{}0$}};
    \draw[black] (A) -- (E);
    \draw[black] (B) -- (E);
    \draw[black] (C) -- (E);
    \draw[black] (D) -- (E);
\end{tikzpicture}
\end{figure}

In the nowcasting exercise our goal is to predict the yearly income distribution using timely information derived from quarterly indicators available during and after the year's end and before the official release in March.  For each year, we compute four nowcasts based on the different information sets given by the information available up to the release of the first, second, third and fourth quarterly data. The perspective is to consider at each update only the information available up to that point in time, so as to reproduce the nowcasts that would have been made in real-time. Our nowcasts origins are therefore in the months of April, July, October and January corresponding respectively to $h = \frac{3}{4} $, $h = \frac{2}{4}$ , $h = \frac{1}{4} $ and $h = 0$. Figure \ref{fig:updates} summarizes the four updates. 


\subsection{Results}
\begin{figure}[H]
    \centering
    \caption{Comparing nowcasts for 1998}\label{fig:example_now}
    \includegraphics[scale=0.2]{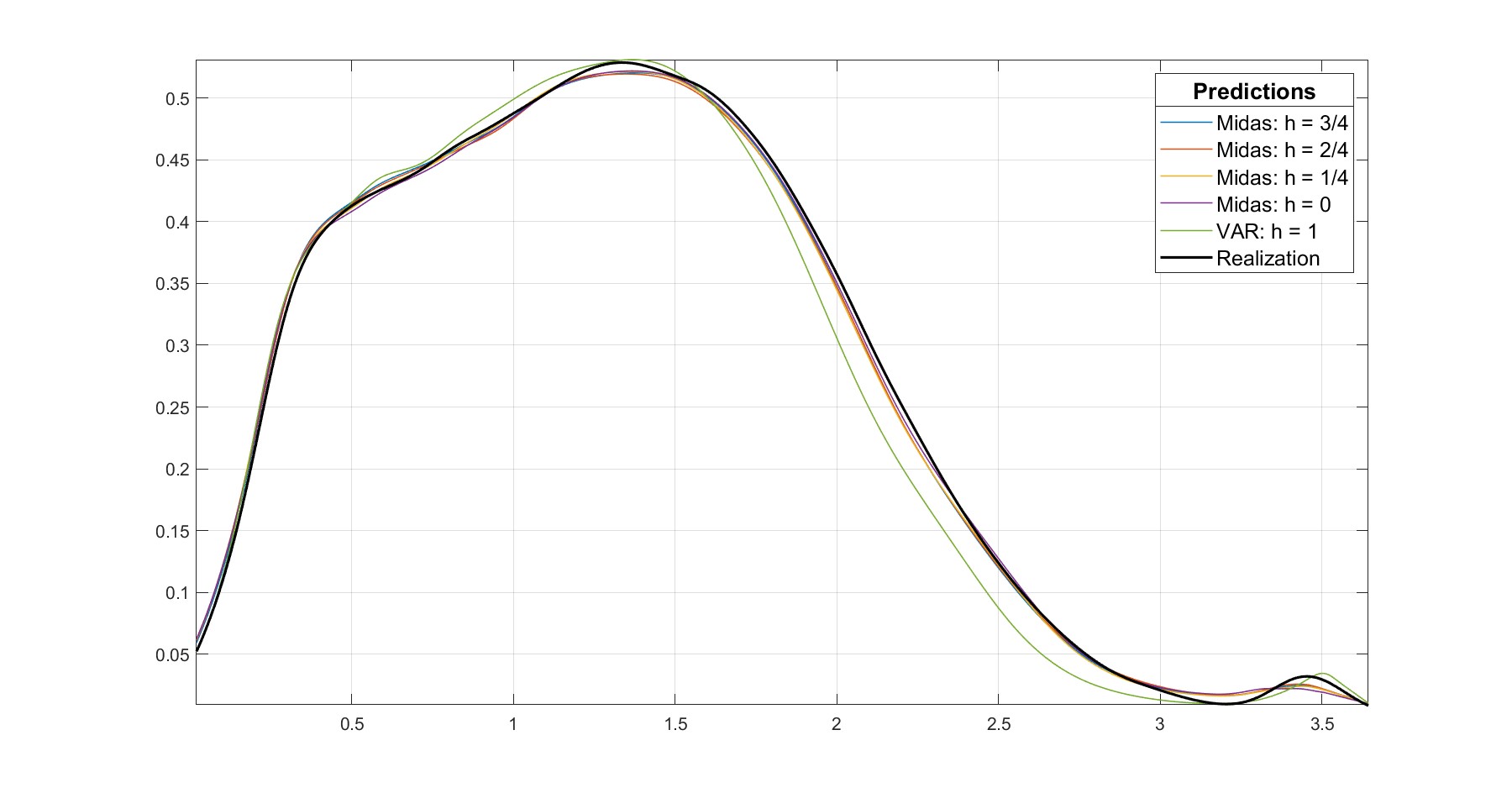}
    \vspace{0.25cm}\hspace{2.3cm}\parbox{1.2\textwidth}{\scriptsize Notes: The figure shows the nowcasts for the hyperbolic-sine transformed distribution \\ of households' income normalized by GDP-per capita for the year 2002. }
\end{figure}

The estimation sample begins in 1968, with predictions for income distribution starting from 1998 to 2023. For each period, we compute nowcasts at horizons $h = \frac{3}{4}$, $h = \frac{2}{4}$, $h = \frac{1}{4}$, and $h = 0$.   Figure \ref{fig:example_now} presents a representative example of our forecasting output for the distribution of income in 1998. The figure compares the predictions generated by the functional MIDAS model in the four updates, alongside one-step-ahead forecasts from the functional VAR model, and the actual observed distribution.\footnote{Also here, the point forecast of the distribution function is obtained by averaging across the posterior distribution draws.}
 As our pseudo-real-time nowcasting exercise is designed to produce nowcasts of the income distribution for each year from 1998 to 2023 in four different updates,  we are reporting the results in four distinct panels in 
 table
 \ref{tab:forecast_results} and \ref{tab:rmse_metrics}. Our competitors in the forecasting race are a simple functional VAR model for the low frequency distribution only and a functional MIDAS model with a ridge-type prior the coefficients in the SUR-MIDAS representation. We consider a ridge type prior both for an unrestricted MIDAS dynamics and for a restricted Almon-lag dynamics.\footnote{Section \ref{sec:ridge_type} in the appendix provides the details on this prior and setting the hyper-parameters in this model.} The  ridge-type prior assigns a distinct degree of shrinkage to the coefficients on the lags of the low frequency factors and the high frequency indicators. While the simple functional VAR model is updated only in April, when the release of the data on the survey of the previous year becomes available, the other models leverage high frequency information within the year to nowcast the low frequency distribution before its release. Overall, the Table shows that the functional MIDAS model does a pretty good job when compared to the simple functional single frequency VAR, with gains across all measures. In particular, including high frequency information allows to increase the accuracy of the forecast for the entire distribution function and some selected moments of this distribution. Notably, including high frequency information allows to reduced the RMSE associated to the two measures of inequality, being the Gini index and the coefficient of variation. This is true in all the updates considered. The spike-and-slab group lasso prior often outperforms the ridge prior in predicting the entire distribution function of income when comparing the KL distance with the true distribution; however, this superiority does not extend to all the metrics considered in this analysis.
\begin{table}[H]
    \centering
    \scalebox{0.7}{
    \begin{tabular}{lccccccc}
        \toprule
        \multicolumn{8}{c}{$h=\frac{3}{4}$} \\ 
        \bottomrule\toprule
        & Avg KL & Avg HD & Avg QS5 & Avg QS20 & Avg QS50 & Avg QS80 & Avg QS95 \\
        \midrule
        FLAT VAR & 0.1612 & 0.0845 & 0.0053 & 0.0114 & 0.0224 & 4407.8732 & 11962.0658 \\
        BVAR & 0.0242 & 0.0441 & 0.0083 & 0.0151 & 0.0187 & 0.0119 & 126.7556 \\
        RIDGE-SUR-unres-MIDAS & 0.0116 & 0.0349 & 0.0069 & 0.0154 & 0.0167 & 0.0110 & 0.0069 \\
        RIDGE-SUR-MIDAS & 0.0114 & 0.0383 & 0.0086 & 0.0171 & 0.0181 & 0.0120 & 0.0086 \\
        BLSS-SUR MIDAS & 0.0090 & 0.0371 & 0.0069 & 0.0096 & 0.0127 & 0.0093 & 0.0058 \\
        \bottomrule
        \toprule
        \multicolumn{8}{c}{$h=\frac{2}{4}$} \\ 
        \bottomrule\toprule
        & Avg KL & Avg HD & Avg QS5 & Avg QS20 & Avg QS50 & Avg QS80 & Avg QS95 \\
        \midrule
        FLAT VAR & 0.1612 & 0.0845 & 0.0053 & 0.0114 & 0.0224 & 4407.8732 & 11962.0658 \\
        BVAR & 0.0242 & 0.0441 & 0.0083 & 0.0151 & 0.0187 & 0.0119 & 126.7556 \\
        RIDGE-SUR-unres-MIDAS & 0.0118 & 0.0349 & 0.0068 & 0.0149 & 0.0162 & 0.0108 & 0.0070 \\
        RIDGE-SUR-MIDAS & 0.0115 & 0.0378 & 0.0086 & 0.0167 & 0.0177 & 0.0117 & 0.0083 \\
        BLSS-SUR MIDAS & 0.0129 & 0.0376 & 0.0064 & 0.0106 & 0.0128 & 0.0088 & 0.0092 \\
        \bottomrule
        \toprule
        \multicolumn{8}{c}{$h=\frac{1}{4}$} \\ 
        \bottomrule\toprule
        & Avg KL & Avg HD & Avg QS5 & Avg QS20 & Avg QS50 & Avg QS80 & Avg QS95 \\
        \midrule
        FLAT VAR & 0.1612 & 0.0845 & 0.0053 & 0.0114 & 0.0224 & 4407.8732 & 11962.0658 \\
        BVAR & 0.0242 & 0.0441 & 0.0083 & 0.0151 & 0.0187 & 0.0119 & 126.7556 \\
        RIDGE-SUR-unres-MIDAS & 0.0117 & 0.0350 & 0.0069 & 0.0152 & 0.0166 & 0.0109 & 0.0069 \\
        RIDGE-SUR-MIDAS & 0.0105 & 0.0374 & 0.0086 & 0.0162 & 0.0172 & 0.0113 & 0.0075 \\
        BLSS-SUR MIDAS & 0.0088 & 0.0374 & 0.0067 & 0.0118 & 0.0148 & 0.0097 & 0.0058 \\
        \bottomrule
        \toprule
        \multicolumn{8}{c}{$h=0$} \\ 
        \bottomrule\toprule
        & Avg KL & Avg HD & Avg QS5 & Avg QS20 & Avg QS50 & Avg QS80 & Avg QS95 \\
        \midrule
        FLAT VAR & 0.1612 & 0.0845 & 0.0053 & 0.0114 & 0.0224 & 4407.8732 & 11962.0658 \\
        BVAR & 0.0242 & 0.0441 & 0.0083 & 0.0151 & 0.0187 & 0.0119 & 126.7556 \\
        RIDGE-SUR-unres-MIDAS & 0.0117 & 0.0347 & 0.0068 & 0.0150 & 0.0163 & 0.0109 & 0.0069 \\
        RIDGE-SUR-MIDAS & 0.0106 & 0.0373 & 0.0084 & 0.0161 & 0.0172 & 0.0114 & 0.0076 \\
        BLSS-SUR MIDAS & 0.0109 & 0.0360 & 0.0066 & 0.0114 & 0.0146 & 0.0102 & 0.0079 \\
        \bottomrule
    \end{tabular}
 }
    \caption{Forecast accuracy from 1998 to 2023}
    \label{tab:forecast_results}
    \vspace{0.25cm}\hspace{0cm}\parbox{1.2\textwidth}{\scriptsize Notes: Please see the text for the definitions of the various evaluation measures. }
\end{table}

\begin{table}[H]
    \centering
    \scalebox{0.7}{
    \begin{tabular}{lccccccc}
        \toprule
        \multicolumn{8}{c}{$h=\frac{3}{4}$} \\ 
        \bottomrule\toprule
        & RMSE Mean & RMSE Variance & RMSE Skewness & RMSE Kurtosis & RMSE IR & RMSE GINI & RMSE CV \\
        \midrule
        FLAT VAR & 0.2932 & 0.0553 & 1.0749 & 3.3231 & 27196.4246 & 0.3993 & 16.1803 \\
        BVAR & 0.0842 & 0.0432 & 0.2978 & 0.6516 & 0.0334 & 0.0677 & 2.3908 \\
        RIDGE-SUR-unres-MIDAS & 0.0387 & 0.0357 & 0.1398 & 0.2913 & 0.0278 & 0.0207 & 1.3575 \\
        RIDGE-SUR-MIDAS & 0.0480 & 0.0394 & 0.1537 & 0.3189 & 0.0274 & 0.0239 & 1.3073 \\
        BLSS-SUR MIDAS & 0.0422 & 0.0388 & 0.1332 & 0.2882 & 0.0330 & 0.0228 & 1.3976 \\
        \bottomrule
        \toprule
        \multicolumn{8}{c}{$h=\frac{2}{4}$} \\ 
        \bottomrule\toprule
        & RMSE Mean & RMSE Variance & RMSE Skewness & RMSE Kurtosis & RMSE IR & RMSE GINI & RMSE CV \\
        \midrule
        FLAT VAR & 0.2932 & 0.0553 & 1.0749 & 3.3231 & 27196.4246 & 0.3993 & 16.1803 \\
        BVAR & 0.0842 & 0.0432 & 0.2978 & 0.6516 & 0.0334 & 0.0677 & 2.3908 \\
        RIDGE-SUR-unres-MIDAS & 0.0381 & 0.0356 & 0.1419 & 0.2953 & 0.0280 & 0.0212 & 1.3591 \\
        RIDGE-SUR-MIDAS & 0.0460 & 0.0386 & 0.1535 & 0.3190 & 0.0277 & 0.0239 & 1.3099 \\
        BLSS-SUR MIDAS & 0.0443 & 0.0372 & 0.1823 & 0.3958 & 0.0314 & 0.0303 & 1.6204 \\
        \bottomrule
        \toprule
        \multicolumn{8}{c}{$h=\frac{1}{4}$} \\ 
        \bottomrule\toprule
        & RMSE Mean & RMSE Variance & RMSE Skewness & RMSE Kurtosis & RMSE IR & RMSE GINI & RMSE CV \\
        \midrule
        FLAT VAR & 0.2932 & 0.0553 & 1.0749 & 3.3231 & 27196.4246 & 0.3993 & 16.1803 \\
        BVAR & 0.0842 & 0.0432 & 0.2978 & 0.6516 & 0.0334 & 0.0677 & 2.3908 \\
        RIDGE-SUR-unres-MIDAS & 0.0389 & 0.0358 & 0.1409 & 0.2932 & 0.0278 & 0.0209 & 1.3618 \\
        RIDGE-SUR-MIDAS & 0.0445 & 0.0385 & 0.1456 & 0.3078 & 0.0278 & 0.0227 & 1.2497 \\
        BLSS-SUR MIDAS & 0.0430 & 0.0390 & 0.1333 & 0.2916 & 0.0295 & 0.0192 & 1.4507 \\
        \bottomrule
        \toprule
        \multicolumn{8}{c}{$h=0$} \\ 
        \bottomrule\toprule
        & RMSE Mean & RMSE Variance & RMSE Skewness & RMSE Kurtosis & RMSE IR & RMSE GINI & RMSE CV \\
        \midrule
        FLAT VAR & 0.2932 & 0.0553 & 1.0749 & 3.3231 & 27196.4246 & 0.3993 & 16.1803 \\
        BVAR & 0.0842 & 0.0432 & 0.2978 & 0.6516 & 0.0334 & 0.0677 & 2.3908 \\
        RIDGE-SUR-unres-MIDAS & 0.0378 & 0.0354 & 0.1406 & 0.2927 & 0.0280 & 0.0209 & 1.3631 \\
        RIDGE-SUR-MIDAS & 0.0443 & 0.0388 & 0.1455 & 0.3057 & 0.0278 & 0.0223 & 1.2314 \\
        BLSS-SUR MIDAS & 0.0420 & 0.0374 & 0.1632 & 0.3243 & 0.0311 & 0.0259 & 1.5719 \\
        \bottomrule
    \end{tabular}
 }
    \caption{Forecast accuracy metrics from 1998 to 2023 RMSE of selected moments}
    \label{tab:rmse_metrics}
    \vspace{0.25cm}\hspace{0cm}\parbox{1.2\textwidth}{\scriptsize Notes: Please see the text for the definitions of the various evaluation measures. }
\end{table}

\begin{figure}[H]
    \centering
    \caption{Inclusion probabilities in the SUR-MIDAS for $h=0$}\label{fig:proh_hist}
    \includegraphics[scale=1.5]{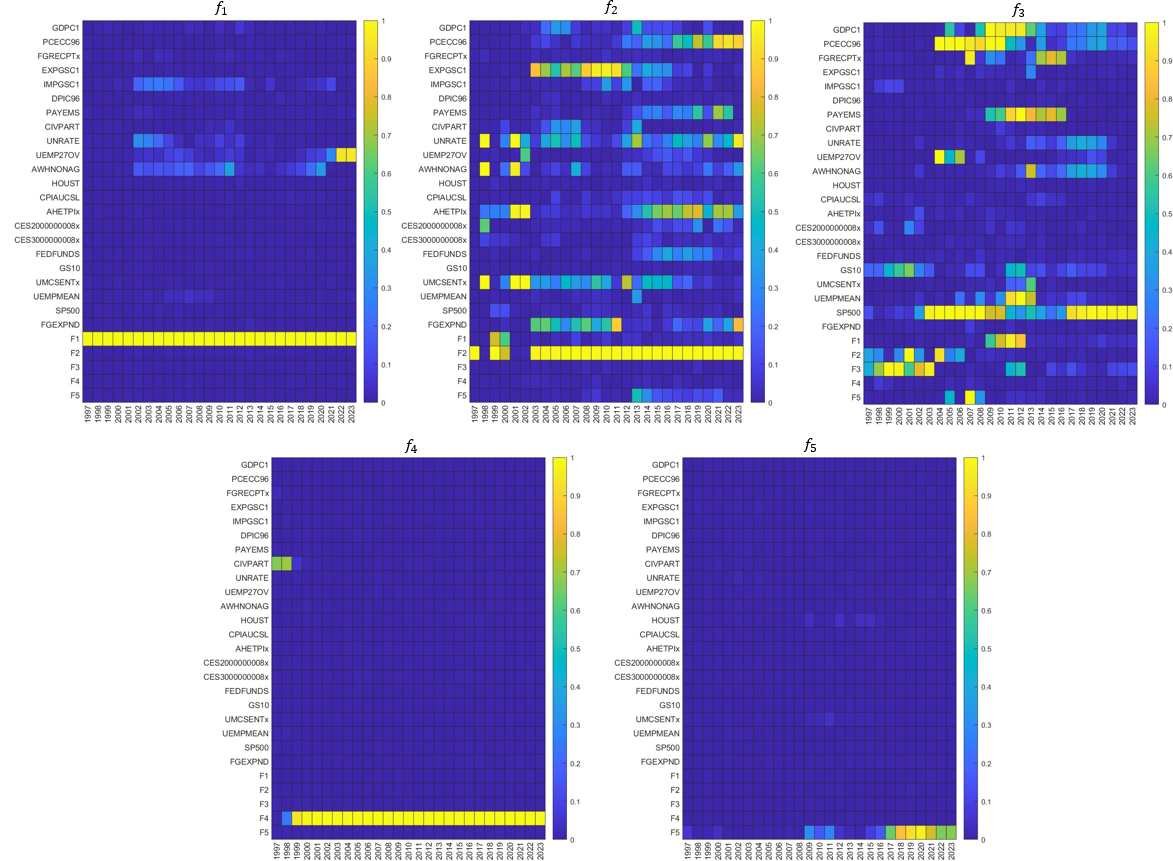}
        \vspace{0.25cm}\hspace{2.3cm}\parbox{1.2\textwidth}{\scriptsize Notes: The heat-map reports the posterior mean probability of inclusion of each group. }
\end{figure}

We can exploit the posterior estimates from the SUR-MIDAS model with spike and slab prior to investigate the set of relevant regressors, and understand whether the relevance of the predictors has changed over time. Figure \ref{fig:proh_hist} reports the posterior mean inclusion probabilities of the regressors in the five equations of the spike and slab lasso SUR-MIDAS model for the factors used for the approximation of the households' income distribution. The model is the one corresponding to $h=0$, with high-frequency information fully matching the low frequency. The figure shows that some predictors have been historically relevant for predicting households' income distribution. More specifically, the lags of the unemployment rate, the number of civilians unemployed for 27 weeks and over, personal consumption expenditure, the number of non-farm employees,  federal consumption expenditure and the SP500 are the most relevant predictors. Some groups, like those consisting of the lags of GDP and exports, were considered relevant and selected in certain periods, but their significance has diminished in recent years. Furthermore, the groups corresponding to the own lags of the first, second and fourth factors are almost always selected and considered relevant, meaning that these factors exhibit a significant degree of persistence. The figure also shows that  the first, fourth and fifth factors are predicted by a small, sparse set of predictors, indicating that only a limited number of predictors contribute significantly to their prediction. In contrast, the second and third factor exhibits less sparsity, meaning that a broader range of predictors are included and contribute to their prediction. 

\section{Conclusions}\label{conclusions}

In this paper, we propose a functional MIDAS model to leverage high-frequency information for nowcasting a distribution observed at a lower frequency. First, we approximate the low-frequency distribution using functional principal component analysis. Then, we specify a spike and slab prior for the parameters of the SUR-MIDAS approximation of the functional MIDAS model. Through a simulation exercise, we demonstrate that this prior is particularly useful for selecting relevant predictors from a potentially large set of high-frequency indicators, even in small samples, as is typically the case when forecasting distributions from yearly household surveys. The simulation also highlights the importance of exploiting high-frequency indicators to accurately predict the entire distribution observed at low frequency, as well as key moments and features that can inform changes in inequality.  

In general, although it does not enable high-frequency predictions for the micro-level distribution, the MIDAS framework is particularly promising for forecasting distributions by leveraging high-frequency information. It allows for imposing less structure when approximating the micro-level variable distribution, which would otherwise be required in the mixed-frequency framework of \cite{schorfheide2015real}. For example, when the observed low-frequency variable of interest is a flow variable that accumulates high-frequency streams, aggregating the unobserved high-frequency distributions is non-trivial and may require parametric assumptions for tractability. 

In the application, we aim nowcasting the March ASEC households' income data from the Current Population Survey leveraging on quarterly macroeconomic indicators available before the release. We find that our model enhances the accuracy of the forecasts for the distribution of households income and of key features of this distribution that signal changes in inequality. This enhanced predictive capability can help policymakers intervene more effectively with policies aimed at tackling inequality and promoting economic stability and growth.

\printbibliography
\clearpage
\appendix
\section{Appendix}

\subsection{SUR-MIDAS approximation of the functional MIDAS model}\label{approx_surmidas}
The functional MIDAS model is given by: 
\begin{equation}  \label{fun_midas_2}
        q_{t}(\tau) = c_q(\tau) + \sum_{l=0}^{p_x-1}\boldsymbol{B}_{qx,l}(\tau)L(l/m)\boldsymbol{x}_t^{(m)}  + \sum_{l=1}^{p_q}\int_0^1 B_{qq,l}(\tau,\tau')L(l)q_{t}(\tau')d\tau' + u_{q,t}(\tau)
\end{equation}
Thanks to the finite dimensional approximation (\ref{finiteapprox}) and assuming $\mu(\tau)=0$ for ease of exposition, we can rewrite the function MIDAS expanding the unknown functions of $\tau$ as follows
\begin{equation}
\begin{aligned}
     c_{q}(\tau) = \boldsymbol{h}_K(\tau)'\tilde{c}_{q,t} \hspace{1cm} & 
    \boldsymbol{B}_{qx,l}(\tau) = \boldsymbol{h}_K(\tau)'\boldsymbol{B_{qx,l}} \hspace{1cm} \\  
    B_{qq,l}(\tau,\tau') = \boldsymbol{h}_K(\tau)'\boldsymbol{B_{qq,l}}\boldsymbol{s}_{K}(\tau') &\hspace{1cm} \boldsymbol{u}(\tau) = \boldsymbol{h}_K(\tau)'\tilde{u}_{q,t}
\end{aligned}
\end{equation}
where $\boldsymbol{s}_{K}(\tau')$ is a $K$-dimensional vector of functional basis such that  
\begin{equation}
\int  \boldsymbol{s}_{K}(\tau) \boldsymbol{h}_K(\tau)'  d \tau= \boldsymbol{C_{f}}
\end{equation}
Plugging in (\ref{fun_midas_2}) we get
\begin{equation}        \boldsymbol{h}_K(\tau)'\boldsymbol{f}_{t;K} =  \boldsymbol{h}_K(\tau)'\tilde{c}_{q,t} + \sum_{l=1}^{p_x} \boldsymbol{h}_K(\tau)'\boldsymbol{B_{qx,l}}L(l/m)\boldsymbol{x}_{t}^{(m)} + \boldsymbol{h}_K(\tau)'\sum_{l=1}^{p_q} \boldsymbol{B_{qq,l}}L(l)\boldsymbol{f}_{t;K} + \boldsymbol{h}_K(\tau)'\tilde{u}_{q,t}        
\end{equation}
that becomes 
\begin{equation}\label{fvar22}
        \boldsymbol{f}_{t;K} =  \tilde{c}_{q,t} + \sum_{l=1}^p \boldsymbol{B_{qx,l}}L(l/m)\boldsymbol{x}_{t}^{(m)} +  \sum_{l=1}^P \boldsymbol{B_{l,qq}}\boldsymbol{C_{f}}L(l)\boldsymbol{f}_{t;K} + \tilde{u}_{q,t}      
\end{equation}
Now define $\boldsymbol{\Phi_{fx,l}}$, $\boldsymbol{\Phi_{ff,l}}$ and $ \boldsymbol{u_{f,t+h}} = \tilde{u}_{q,t}$ and we obtain the SUR MIDAS model

\begin{equation}\label{surmidas2}
    \boldsymbol{f}_{t;K} = \boldsymbol{\phi}_{0} + \sum_{l=1}^{p_x} \boldsymbol{\Phi_{fx,l}}L(l/m)\boldsymbol{x}_t^{(m)}  + \sum_{l = 1}^{p_q}\boldsymbol{\Phi_{ff,l}}L(l)\boldsymbol{f}_{t;K} + \boldsymbol{u_{f,t}}
\end{equation}

\clearpage
\subsection{Conditional posterior distributions in the Gibbs Sampler}\label{condpost}

Exploiting the direct representation of the Almon lag polynomial we can rewrite the SUR-MIDAS model as

\begin{equation}
\begin{aligned}
    f_{1t} & = \boldsymbol{z}_t'\boldsymbol{\theta}_1 + \varepsilon_{1t} \hspace{5cm}  \varepsilon_{1t} \sim \mathcal{N}(0, \sigma^2_1) \\
    \vdots \\
    f_{2t} & = \boldsymbol{z}_t'\boldsymbol{\theta}_2+ \alpha_{21} u_{1t} + \varepsilon_{2t}  \hspace{3.5cm}  \varepsilon_{2t} \sim \mathcal{N}(0, \sigma^2_2) \\
    \vdots \\
    f_{Kt} & = \boldsymbol{z}_t'\boldsymbol{\theta}_K + \alpha_{K,1} u_{1t} + \ldots + \alpha_{K,K-1} u_{K-1t} + \varepsilon_{Kt}  \hspace{2cm}  \varepsilon_{Kt} \sim \mathcal{N}(0, \sigma^2_K) \\
\end{aligned}
\end{equation} 
where the vector $\boldsymbol{\theta}_i = [\boldsymbol{\theta_{i,1}}', \ldots, \boldsymbol{\theta_{i,G}}' ]'$ collects all the parameters of the Almon lag polynomial for all the  $G = (n_x + K)$ groups in the $i^{th}$ equation for $i = 1, \ldots, K$. Note that each vector $\boldsymbol{\theta}_{ij}$ is of dimension $(p_a + 1 - r_a) \times 1$. Define $\boldsymbol{F} $ as the $T \times K$ matrix with the $T \times 1$ vector $\boldsymbol{f}_i = [f_{i1},\ldots,f_{iT}]'$  stacked on its columns and define $\boldsymbol{\Theta}$ the $G(p_a + 1 - r_a) \times K$ matrix stacking the $\boldsymbol{\theta}_i$ on its columns. 
In compact form the model is
\begin{equation}
    (\boldsymbol{F} - \boldsymbol{Z}\boldsymbol{\Theta})\boldsymbol{\tilde{A}'} = \boldsymbol{\varepsilon}
\end{equation}
where $\boldsymbol{\tilde{A}}$ is the lower triangular matrix with unit elements on the main diagonal and rows given by  $\boldsymbol{\tilde{\alpha}_{i,.}} = [-\alpha_{i,1}, \ldots, -\alpha_{i i-1}]$. Exploiting the triangularization in \citet{corrCARRIERO2022} to rewrite the equations that involve the coefficients in the $i^{th}$ equation i.e. $\boldsymbol{\theta}_i$  as
\begin{equation}
  vec((\boldsymbol{F} - \boldsymbol{Z \Theta}^{[i = 0]} ) \boldsymbol{\tilde{A}}^{(i:K, 1:K)'}) = \left( \boldsymbol{\tilde{A}}^{(i:K, i)} \otimes \boldsymbol{Z} \right) \boldsymbol{\theta}_{i} + vec\left( \varepsilon^{(i:K)} \right) 
\end{equation}
where the notation $\boldsymbol{\Theta}^{[i = 0 ]}$ denotes the matrix $\boldsymbol{\Theta}$ with the $i^{th}$ column set equal to zero, while the notation $\boldsymbol{\tilde{A}}^{(i:K, i)}$ denotes the elements from row  $i$ to row $K$ in the $i^{th}$ column of $\boldsymbol{\tilde{A}}$. Define
\begin{equation}
\boldsymbol{\tilde{Z}} = (\boldsymbol{\tilde{A}}^{(j:K,j)} \otimes \boldsymbol{Z})
 \end{equation} 
and $\boldsymbol{\tilde{Z}}_j $ the $j^{th}$ column of $\boldsymbol{\tilde{Z}}$ while  $\boldsymbol{\tilde{Z}}_{\backslash j}$ the matrix $\boldsymbol{\tilde{Z}}$ without the $j^{th}$ column. Now rewrite it as
\begin{equation}
  vec((\boldsymbol{F} - \boldsymbol{Z \Theta}^{[i = 0]} ) \boldsymbol{\tilde{A}}^{(i:K, 1:K)'})  = \boldsymbol{\tilde{Z}_j} \boldsymbol{\theta}_{ij} + \boldsymbol{\tilde{Z}}_{\backslash j}\boldsymbol{\theta}_{i\backslash j} + vec\left( \boldsymbol{\varepsilon}^{(i:K)} \right)
\end{equation} 
then consider
\begin{equation}
  vec((\boldsymbol{F} - \boldsymbol{Z \Theta}^{[i = 0]} ) \boldsymbol{\tilde{A}}^{(i:K, 1:K)'})  - \boldsymbol{\tilde{Z}}_{\backslash j}\boldsymbol{\theta}_{i\backslash j} = \boldsymbol{\tilde{Z}_j} \boldsymbol{\theta}_{ij} +  vec\left( \boldsymbol{\varepsilon}^{(i:K)} \right)
\end{equation} 
Defining $\boldsymbol{\tilde{f}}_{ij} = vec((\boldsymbol{F} - \boldsymbol{Z \Theta}^{[i = 0]} ) \boldsymbol{\tilde{A}}^{(i:K, 1:K)'})  - \boldsymbol{\tilde{Z}}_{\backslash j}\boldsymbol{\theta}_{i\backslash j}$ becomes
\begin{equation}
\boldsymbol{\tilde{f}}_{ij}= \boldsymbol{\tilde{Z}_j} \boldsymbol{\theta}_{ij} +  vec\left( \boldsymbol{\varepsilon}^{(i:K)} \right)
\end{equation}
To obtain the conditional posterior of the $\boldsymbol{\theta_{ij}}$'s just combine this equation with the prior for $\boldsymbol{\theta}_{ij}$ in equation (\ref{PRIOR_THETA}). In particular we get 
\begin{equation}
    \begin{aligned}
      \boldsymbol{\theta}_{ij} \mid \boldsymbol{\theta}_{i \backslash j}, \boldsymbol{\theta}_{1:i-1,}, \tau_{ij}, \lambda, \pi_{1i}, \pi_{0i}, \boldsymbol{f}, \boldsymbol{Z} & \sim (1 - \pi_{1ij}) \mathcal{N} \left( \boldsymbol{P}_{ij}^{-1} \boldsymbol{C}_{ij}, \sigma_i^2 \boldsymbol{P}_{ij}^{-1} \right) + \pi_{1ij} \delta_0 (\boldsymbol{\theta}_{ij}) \\ 
\boldsymbol{P}_{ij} =   \tau^2_{ij} \boldsymbol{I}_{gj} + \boldsymbol{\tilde{Z}}_j' \boldsymbol{\Sigma}^{\left(i:K\right)^{-1}} \boldsymbol{\tilde{Z}}_j'
\\
\boldsymbol{C}_{ij} = \boldsymbol{\tilde{Z}}_j' \boldsymbol{\Sigma}^{\left(i:K\right)^{-1}} \boldsymbol{\tilde{f}}_{ij}
    \end{aligned}
\end{equation}
To write the conditional posterior of the elements in $\boldsymbol{\tilde{A}}$ we define  
\begin{equation}
    \boldsymbol{u_{i}} = \boldsymbol{f_{i}}  -  \boldsymbol{Z}\boldsymbol{\theta}_i
\end{equation}
where $\boldsymbol{f_{i}}$ is $T \times 1$ 
and $\boldsymbol{Z}$ is the $T \times (p_a + 1 - r_a)G$ design matrix.  Furthermore for $i=2,\ldots, K$ define $\boldsymbol{S}_i = diag(s^2_{1}, \ldots,s^2_{i})$. Define also 
\begin{equation}        \boldsymbol{\tilde{U}_{i}} = [\boldsymbol{u}_1, \ldots,    \boldsymbol{u}_{i-1}]' \hspace{2cm} \text{for $2=1, ,K$}
\end{equation}
and $\boldsymbol{A}$ the lower triangular matrix with unit elements on the main diagonal, such that the free elements in each row $i$ are given by $\boldsymbol{\alpha_{i,.}} = [\alpha_{i,1}, \ldots, \alpha_{i i-1}]$ for $i=1, \ldots, K$.

\begin{equation}
    \begin{aligned}
       \boldsymbol{\alpha_{i,.}}' \sim \mathcal{N} \left( \left(\boldsymbol{\tilde{U}_i}'\boldsymbol{\tilde{U}_i} + \boldsymbol{S}_i^{-1} \right)^{-1} (\boldsymbol{\tilde{U}_i}'\boldsymbol{u_{i}}) , \left(\boldsymbol{\tilde{U}_i}'\boldsymbol{\tilde{U}_i} + \boldsymbol{S}_i^{-1} \right)^{-1} \sigma^2_i \right) \hspace{1cm} \text{for $2=1, ,K$}\\  
    \end{aligned}
\end{equation}
The other conditional posteriors are given by

\begin{equation}
    \resizebox{0.8\hsize}{!}{ 
        $
        \begin{aligned}
            \sigma_i^2 \mid \boldsymbol{\theta}, \tau_{ij}, \lambda_{ij}, \pi_{1i}, \pi_{0i}, \boldsymbol{f}, \mathbf{Z} & \sim \text{Inv-Gamma} \left( \frac{T + \tilde{G}_i - 1}{2} + \frac{(i-1)}{2} +  \frac{(v_{0i} + i - K)}{2}, \right. \\
            & \quad \left. \frac{1}{2} \|(\boldsymbol{Y} - \boldsymbol{Z}\boldsymbol{\Theta} )\boldsymbol{\tilde{A}'} \|_2^2 + \frac{1}{2} \sum_{j=1}^{G} \frac{\| \boldsymbol{\theta}_{ij} \|_2^2}{\tau_{ij}^2} + \frac{s_i^2}{2} \right)
        \end{aligned}
        $
    }
\end{equation}

\begin{equation}
    \begin{aligned}
        \tau_{ij}^{-2} \mid \boldsymbol{\theta}, \sigma_i^2, \lambda_{ij}, \pi_{1i}, \pi_{0i}, \boldsymbol{f}, \mathbf{Z} & \sim 
\begin{cases}
\text{Inv-Gaussian} \left( \frac{\lambda_{ij} \sigma_i}{\|\boldsymbol{\theta}_{ij}\|_2}, \lambda_{ij}^2 \right) & \text{if } \mathbf{\theta}_{ij} \neq 0 \\
\text{Gamma} \left( \frac{g_{ij} + 1}{2}, \frac{\lambda_{ij}^2}{2} \right) & \text{if } \mathbf{\theta}_{ij} = 0
\end{cases} \\
    \end{aligned}
\end{equation}
\begin{equation}
    \begin{aligned}
        \pi_{0i} \mid \boldsymbol{\theta}, \sigma_i^2, \tau_{ij}, \lambda, \pi_{1i}, \boldsymbol{f}, \mathbf{Z} & \sim \text{Beta} \left( \sum_{j=1}^{G} (1 - \gamma_{ij}) + c, \sum_{j=1}^{G} \gamma_{ij} + d \right)
    \end{aligned}
\end{equation}
where $v_{i0}$ is set equal to $ 1 + \frac{i}{2}$, $\tilde{G}_i = \sum_{j=1}^G g_{ij}\gamma_{ij}$, $d =1$, $c = \bar{\kappa}G^{\nu}$, $\bar{\kappa} = (1  + G^{-1})$ and 
\begin{equation}
\gamma_{ij} =
   \begin{cases} 
1 & \text{if } \boldsymbol{\theta_{ij}} \neq 0 \\ 
0 & \text{if } \boldsymbol{\theta_{ij}} = 0 
\end{cases} 
\end{equation}
\begin{equation}
    \pi_{1,ij} = P(\boldsymbol{\theta_{ij}} = 0 | \theta_{\setminus j}, \sigma^2, \tau, \lambda, \pi_{0i},\boldsymbol{f}, \mathbf{Z}) = \frac{\pi_{0i}}{\pi_{0i} + (1 - \pi_{0i}) \left[ (\tau_j^2)^{-\frac{g_j}{2}} |\mathbf{P}_{ij}|^{-\frac{1}{2}} \exp\left( \frac{1}{2\sigma^2} \mathbf{C}_{ij}' \mathbf{P}_{ij}^{-1} \mathbf{C}_{ij} \right) \right]}
\end{equation}
Assuming a conjugate Gamma prior on the penalty hyper-parameters \( \lambda^2_{ij} \) parameterized by shape parameter \( a_2 \) and rate parameter \( b_2 \) we get the following conditional posterior
\begin{equation}
    \begin{aligned}
\lambda_{ij}^2 \mid \boldsymbol{\theta}, \sigma_i^2, \tau_{ij}, \pi_{1i}, \pi_{0i}, \boldsymbol{f}, \mathbf{Z} & \sim \text{Gamma} \left( \frac{g_{ij} + 1}{2} + a_2, \frac{\tau_{ij}^2}{2} + b_2 \right) \\ 
    \end{aligned}
\end{equation}

\subsection{Monte Carlo simulation}\label{sec:appendix_monte_carlo}
Here we report the details on the parameters of the Monte Carlo exercise in section \ref{monte_carlo}. We mostly built on the Monte Carlo exercise in \citet{MOGLIANI2021833} for what concerns setting the value of the parameters of the DGP, extending it to a SUR framework. More in the specific, we set  $\boldsymbol{\Sigma_\epsilon}$  such that the correlation between $x_{j,t}^{(m)}$ and $x_{j',t}^{(m)}$ with $j' \neq j$ is 0.50. In equation (\ref{dyn_HF}) we set $\mu = 0.1$ and $\rho = 0.5$. In equation (\ref{dyn_LLF}) and we set $\alpha_i = 0$ and
\begin{equation}
    \resizebox{0.8\hsize}{!}{ 
        $
        \boldsymbol{\beta_1} = 
        \begin{pmatrix}
        0 \\ 0.3 \\ 0.3 \\  0 \\ -0.5 \\ 0.1 \\ 0  \\ 0 \\ 0.3 \\ \boldsymbol{0}_{21 \times 1} 
        \end{pmatrix}  
        \hspace{0.5cm}
        \boldsymbol{\beta_2} = 
        \begin{pmatrix}
         \boldsymbol{0}_{19 \times 1} \\ 0.3 \\ 0.3 \\  0.1 \\ 0 \\ -0.5 \\ -0.3  \\ 0 \\ 0.3 \\ 0.1 \\  \boldsymbol{0}_{2 \times 1}
        \end{pmatrix} 
        \hspace{0.5cm}
        \boldsymbol{\beta_3} = 
        \begin{pmatrix}
        0 \\ 0.3 \\ 0.5 \\  0 \\ -0.1 \\ \boldsymbol{0}_{21 \times 1} \\ 0.3  \\ 0 \\ 0 \\ 0.5  
        \end{pmatrix} 
        \hspace{0.5cm}
        \boldsymbol{\Phi_1} = 
        \begin{bmatrix} 
        0.8 & 0.001 & 0 \\
        0.02 & 0.5 & 0 \\
        -0.01 & 0 & 0.5
        \end{bmatrix}
        \hspace{0.5cm}
        \boldsymbol{\Phi_2} =
        \begin{bmatrix} 
        0.02 & 0 & 0.1 \\
        0 & 0.03 & 0.04 \\
        0 & -0.02 & 0.05
        \end{bmatrix}
        $
    }
\end{equation}
Conditional on these parameters, we set the diagonal elements of $\boldsymbol{\Omega}$ such that the noise-to-signal ratio of the mixed-frequency regression is 0.20. The off diagonal elements of $\boldsymbol{\Omega}$ are instead set to imply $corr(u_{1,t},u_{2,t}) = 0.1$, $corr(u_{1,t},u_{3,t}) = -0.1$ and $corr(u_{2,t},u_{3,t}) = 0.2$.

\subsection{Ridge type shrinkage prior for the unrestricted SUR-MIDAS}\label{sec:ridge_type}
We compare the lasso-type shrinkage prior with a ridge-type shrinkage prior for the coefficients in the SUR-MIDAS model. This prior is inspired by the shrinkage prior for a univariate unrestricted MIDAS regression in  \citet{CCMJRSS}.  More specifically, we rewrite the SUR-MIDAS model in compact form as 
\begin{equation}
    \boldsymbol{F} = \boldsymbol{X}\boldsymbol{\Phi} + \boldsymbol{U}
\end{equation}
where $\boldsymbol{F}$ is the $T \times K$ matrix of factors, $\boldsymbol{X}$ is the $T \times ( n_xp_{x} + Kp_{q} )$ design matrix of regressors, which contains both the lags of the high frequency macroeconomic indicators and the lags of the low frequency factors.  $\boldsymbol{U}$ is a $T \times N$ matrix of correlated errors such that $E[\boldsymbol{U'U}]= \boldsymbol{\Omega}$. We specify the following prior for $(\boldsymbol{\Phi},\boldsymbol{\Omega})$
\begin{equation}
vec(\boldsymbol{\Phi})|\boldsymbol{\Omega} \sim \mathcal{N} \left(\underline{\boldsymbol{\Phi}}, \boldsymbol{\underline{V}} \right) 
\end{equation}
\begin{equation}
\boldsymbol{\Omega}\sim \mathcal{IW} \left(\boldsymbol{S}, v_0\right) 
\end{equation}
where $\boldsymbol{\underline{\Phi}}$ is centering the coefficient of the first own lag of the low frequency factors on $0.8$ and all the other coefficients to 0. Then the variance covariance matrix of the prior distribution is given by
\begin{equation}
\boldsymbol{\underline{V}} 
 = \boldsymbol{\Omega} \otimes \boldsymbol{\underline{V_0}}    
\end{equation}
where the elements of the diagonal matrix $\boldsymbol{\underline{V_0}}$ are 
\begin{equation}
\omega_{i} = 
\begin{cases}
 \theta_1 & \text{intercept}  \\
    \frac{\theta_2}{s^2_r l^{\theta^3}} & \text{on the lags of the low frequency factors}  \\
    \frac{\theta_4}{s^2_r l^{\theta^5}}       & \text{on the lags of the high frequency variables}  \\
\end{cases}
\end{equation}
The shrinkage hyper-parameters $[\theta_1,\theta_2, \theta_3, \theta_4, \theta_5]$
are selected to maximize the marginal data density of the model which is available in close form. We resort to automatic differentiation to perform the maximization of the marginal likelihood with respect to these hyper-parameters following \citet{chan2020efficientAD}. For the model with the restricted Almon-lag dynamics we consider the same prior with $l=1$.




\clearpage

\end{document}